\documentclass[runningheads]{llncs}
\usepackage[T1]{fontenc}
\usepackage{graphicx}
\usepackage{tikz}
\usepackage{booktabs}
\usepackage{subcaption}
\usepackage{amsmath}
\usepackage{float}
\usepackage{array}
\usepackage{listings}
\usetikzlibrary{positioning}
\begin{document}
\title{Hybrid Voxel Formats for Efficient Ray Tracing}
%
%
\author{Russel Arbore\inst{1} \and
Jeffrey Liu\inst{1} \and
Aidan Wefel\inst{1} \and
Steven Gao\inst{1} \and
Eric Shaffer\inst{1}}
\authorrunning{R. Arbore et al.}
%
\institute{University of Illinois Urbana-Champaign, Urbana IL 61801, USA\\
\email{\{rarbore2,jliu179,awefel2,hongyig3,shaffer1\}@illinois.edu}}
\maketitle              
\begin{abstract}
Voxels are a geometric representation used for rendering volumes, multi-resolution models, and indirect lighting effects. Since the memory consumption of uncompressed voxel volumes scales cubically with resolution, past works have introduced data structures for exploiting spatial sparsity and homogeneity to compress volumes and accelerate ray tracing. However, these works don’t systematically evaluate the trade-off between compression and ray intersection performance for a variety of storage formats. We show that a hierarchical combination of voxel formats can achieve Pareto optimal trade-offs between memory consumption and rendering speed. We present a formulation of "hybrid" voxel formats, where each level of a hierarchical format can have a different structure. For evaluation, we implement a metaprogramming system to automatically generate construction and ray intersection code for arbitrary hybrid formats. We also identify transformations on these formats that can improve compression and rendering performance. We evaluate this system with several models and hybrid formats, demonstrating that compared to standalone base formats, hybrid formats achieve a new Pareto frontier in ray intersection performance and storage cost.

\end{abstract}

\section{Introduction}

Voxels are a type of geometric primitive that explicitly represent volumes as grids of discrete cubes with associated data. Volume representations enable several applications, including volumetric rendering \cite{Gigavoxels,Volumetric1}, medical visualization \cite{MedicalVoxels}, and fluid simulation \cite{FluidSim1}. However, the use of voxels has been limited by the large amount of memory required to store volumetric data at high resolutions.

Several hierarchical formats have been proposed to tackle this problem \cite{EfficientSparseVoxelOctrees,SVDAGs,SSVDAGs,Gigavoxels,VDB,GVDB,NanoVDB,Taichi,Brickmap,Herzberger2023ResidencyOctree}. These formats identify regions of homogeneous voxels and cull empty regions, providing significant compression while serving as ray tracing acceleration structures. However, there is no single best format that simultaneously minimizes both storage size and rendering cost. Our key observation is that each voxel storage format presents specific trade-offs between compression and rendering performance. However, to the best of our knowledge, there has been no systematic study that effectively explores this search space.

We present a formulation for this search space by hierarchically composing multiple "base" formats to form "hybrid" voxel formats. We discuss the implementation of construction and rendering algorithms for arbitrary hybrid voxel formats. We present an evaluation that illuminates many of the trade-offs involved in choosing a format for storing and rendering a particular volume.

Our primary contributions are:
\begin{itemize}
    \item A general compositional formulation of hybrid voxel formats for rapid exploration of complex voxel formats (Section~\ref{hybridformats}).
    \item A metaprogramming system for automatically generating construction and intersection code for arbitrary hybrid voxel formats, including transformations that present optimization opportunities (Section~\ref{metaprogrammingsystem}).
    \item An out-of-core CPU construction algorithm for arbitrary hybrid voxel formats based on Morton order, which generalizes the SVO construction algorithm presented by \cite{OutOfCoreConstructionVoxel} (Section~\ref{oooconstruction}).
    \item An evaluation of a large number of hybrid voxel formats, demonstrating that such formats are able to achieve Pareto optimal memory-performance profiles compared to standalone base formats (Section~\ref{evaluation}).
\end{itemize}

\section{Related Work}

We focus on lossless storage formats for voxel volumes and briefly discuss algorithms for converting triangle models into voxel volumes.

Amanatides and Woo \cite{FastVoxelRayTracing} describe the first modern work on voxel ray tracing---they ray march through a uniform grid of voxels by finding the minimum distance needed to reach the next voxel at each step. Due to the algorithm's simplicity and the regular data layout of raw grids, this method can ray march through lower resolutions models quickly. However, this structure requires occupancy information for each individual voxel, so memory usage scales cubically with resolution. Furthermore, at higher resolutions, the algorithm may spend many iterations marching through empty voxels before finding an intersection.

Laine and Karras \cite{EfficientSparseVoxelOctrees} address the memory usage and intersection scalability of raw voxel grids with Sparse Voxel Octrees (SVOs), where voxel volumes are recursively subdivided into 8 child nodes. Nodes that do not contain any voxels can be culled---this saves memory and allows for rays to jump over large empty spaces. However, the intersection code becomes more complex---the code must perform a pre-order traversal of the octree, requiring a per-thread stack.

K\"{a}mpe et al. \cite{SVDAGs} observe that some non-empty regions are identical across SVO nodes. Identical SVO nodes can be de-duplicated, producing Sparse Voxel Directed Acyclic Graphs (SVDAGs). Villanueva et al. \cite{SSVDAGs} propose de-duplicating nodes that are equivalent under a symmetric transform. They achieve further compression while sacrificing some ray-intersection performance.

Crassin et al. \cite{Gigavoxels} generalize SVOs as $N^3$-trees. The choice of $N$ affects the breadth and depth of the tree, which affects rendering performance and memory usage. SVO nodes are streamed onto the GPU, so that models larger than GPU memory can be rendered interactively. Wingerden \cite{Brickmap} also presents a streaming approach for rendering, but uses a fixed hierarchy of two raw grids.

Previous works have proposed the VDB data structure, which is a hierarchical format similar to B+ trees \cite{VDB,GVDB,NanoVDB}. This format is similar to our hybrid format formulation, because each level of the VDB structure is configured with a separate resolution. This enables memory-performance exploitation \textit{within the domain of VDB formats}. Yuanming et al. \cite{Taichi} present a language for describing computations on spatial data structures. They focus on nested raw grids for dense volumes and hash maps for unbounded volumes. To the best of our knowledge, no previous work has presented how different voxel formats can be combined to form a format space or a systematic evaluation of many such formats.

A common method of obtaining voxel models is voxelizing a triangle mesh \cite{VoxelizePoly,HardwareVoxelization,ParallelVoxelization,FastSceneVoxelization}. To support higher-resolution models, Baert et al. \cite{OutOfCoreConstructionVoxel} present an out-of-core algorithm to construct SVOs for volumes which can't fit in main memory. By iterating over voxels in Morton order, SVOs can be constructed in a bottom-up manner. P\"{a}tzold and Kolb \cite{OOOSVOSGPU} extend this method to support out-of-core voxelization on the GPU.

\section{Hybrid Voxel Formats} \label{hybridformats}

We introduce a generalization of previously proposed voxel formats, called "hybrid" voxel formats. A hybrid format consists of a list of "base" formats, called "levels". These base formats are schemes for storing volumetric data as described in previous work. Base formats are composed to build hybrid formats.

In our implementation, each rendered voxel stores a red, green, blue, and opacity value. A voxel is "empty" if all of its 32 bits are 0 - this implies its opacity is 0, meaning it has no impact on the rendering of the volume. Although we make these assumptions in our experiments, none of the techniques we describe preclude storing more or fewer bits per voxel.

\subsection{The Base Formats}

We use four base formats: uniform voxel grids (\textbf{Raw}), distance fields (\textbf{DF}), sparse voxel octrees (\textbf{SVO}), and sparse voxel directed acyclic graphs (\textbf{SVDAG}). \textbf{Raw} stores voxel data as one array, providing no compression or intersection acceleration. \textbf{DF} also stores the L1 norm distance to the nearest non-empty voxel per voxel, which is used to accelerate ray intersection. We formulate this as a separate format because the L1 distance is additional information used to accelerate rendering while not affecting the semantic content of the underlying volume. \textbf{SVO} uses SVOs to prune large empty spaces in a volume. This also accelerates ray intersection, as ray steps can skip over large chunks of empty space \cite{EfficientSparseVoxelOctrees}. \textbf{SVDAG} is a format similar to \textbf{SVO}, except that nodes are de-duplicated to increase compression \cite{SVDAGs}.

These base formats have trade-offs---\textbf{DF}s provide acceleration for ray marching but have high storage cost. \textbf{SVDAG}s provide excellent compression but less acceleration for ray marching. Additionally, while \textbf{SVDAG}s can effectively compress large voxel volumes, this requires higher per-node overhead. For models where compression via de-duplication is ineffective, either because they are small or are highly heterogeneous, \textbf{SVO}s may provide better compression.

\subsection{Composing Formats}\label{composition}

Hybrid formats are built by composing base formats, which are listed in Table~\ref{tab:formats}. We call a voxel in an "upper" level format a "sub-volume", and a voxel in the lowest level format a "single voxel"---the geometry inside a sub-volume at some level is stored in the sub-volumes of the next level. Previous work applies this idea to voxel grids \cite{VDB,Taichi}. An example of such a format is R(1, 0, 2) R(2, 2, 2)---each voxel in the upper level $2 \times 1 \times 4$ grid points to a sub-volume grid of size $4 \times 4 \times 4$. We extend this idea to support \textbf{DF}, \textbf{SVO}, and \textbf{SVDAG} levels as well. For example, R(1, 0, 2) D(2, 2, 2, 4) is a format consisting of a \textbf{Raw} grid, where each voxel contains a pointer to a \textbf{DF} volume. 


\begin{table}
  \centering
  \begin{tabular}{c c c}
    \toprule
    Format & Signature & Description\\
    \midrule
    \textbf{Raw} & R(W, H, D) & Raw grid of $({2^\text{W}} \times {2^\text{H}} \times {2^\text{D}})$ voxels \\
    \textbf{DF} & D(W, H, D, M) & $({2^\text{W}} \times {2^\text{H}} \times {2^\text{D}})$ voxel grid with max L1 distance M \\
    \textbf{SVO} & S(L) & Sparse Voxel Octree with max depth L\\
    \textbf{SVDAG} & G(L) & Sparse Voxel Directed Acyclic Graph with max depth L\\
  \bottomrule
  \end{tabular}
  \caption{Description of Implemented Formats.}
  \label{tab:formats}
\end{table}

\subsection{Memory Layout} \label{memlayout}

We store all hybrid voxel formats as word-addressable buffers of 32-bit integers. The largest offset that can be stored in one word is $2^{32}$, so the largest volume size our implementation supports is 16 GiB---this limitation could be addressed by using multiple buffers with indirection. We define a "terminating integer" in a voxel format as either a pointer to a sub-volume in the next level's format or an RGBA color for single voxels.

\begin{figure}[ht]
\centering

\begin{minipage}[t]{.50\linewidth}
\begin{lstlisting}[language=C++]
union TermInt {
    uint32_t offset;
    uint32_t rgba;
};
struct Raw {
    TermInt voxels[];
};
struct DF {
    struct DFEntry {
        TermInt voxel;
        uint32_t l1_dist;
    } voxels[];
};
struct SVMasks {
    uint8_t valid_mask;
    uint8_t leaf_mask;
};
\end{lstlisting}
\end{minipage}%
\begin{minipage}[t]{.50\linewidth}
\begin{lstlisting}[language=C++]
struct SVONode {
    union SVOPointer {
        uint32_t child_svo;
        TermInt leaf;
    } children;
    SVMasks masks;
};
struct SVDAGNode {
    union SVDAGFirst {
        SVMasks masks;
        TermInt leaf;
    } first;
    uint32_t children[];
};
\end{lstlisting}
\end{minipage}

\caption{Pseudo-C++ types describing the memory layout of base formats.}
\label{fig:layout}
\end{figure}

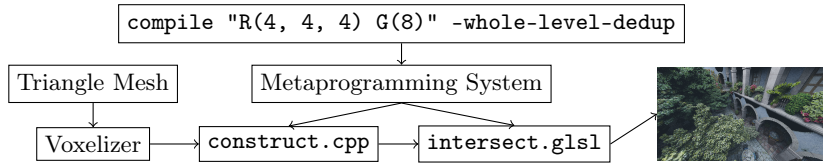
\begin{figure}
\centering

\begin{tikzpicture}

\node[rectangle,draw] at (0cm, 0cm) (a) {\texttt{compile "R(4, 4, 4) G(8)" --whole-level-dedup}};
\node[rectangle,draw] at (0cm, -0.8cm) (b) {Metaprogramming System};
\node[rectangle,draw] at (-1.5cm, -1.6cm) (c) {\texttt{construct.cpp}};
\node[rectangle,draw] at (1.5cm, -1.6cm) (d) {\texttt{intersect.glsl}};
\node[rectangle,draw] at (-4.1cm, -0.8cm) (e) {Triangle Mesh};
\node[rectangle,draw] at (-4.1cm, -1.6cm) (f) {Voxelizer};
\node[inner sep=0pt] at (4.5cm, -1.2cm) (g) {\includegraphics[width=0.18\textwidth]{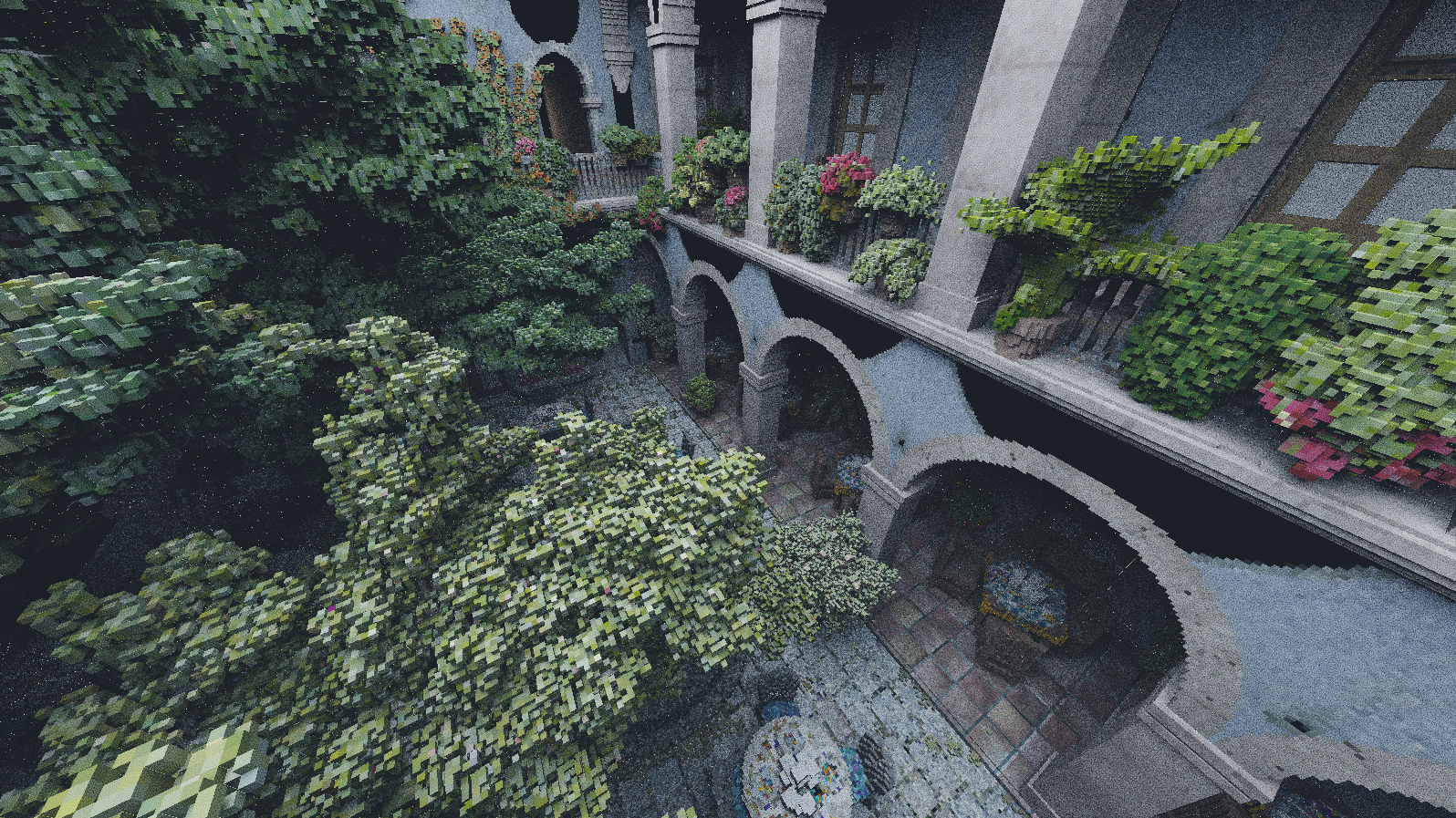}};

\draw[->] (a.south) -- (b.north);
\draw[->] (b.south) -- (c.north);
\draw[->] (b.south) -- (d.north);
\draw[->] (c.east) -- (d.west);
\draw[->] (e.south) -- (f.north);
\draw[->] (f.east) -- (c.west);
\draw[->] (d.east) -- (g.west);

\end{tikzpicture}

\caption{An example usage of the metaprogramming system to compile the \texttt{R(4, 4, 4) G(8)} format with whole level de-duplication.}
\label{fig:metaprogramming_system}
\end{figure}

We store \textbf{Raw} sub-volumes as arrays of integers. Each integer is a terminating integer. If all of the voxels in a subvolume at the next level are empty, the corresponding voxel pointer is 0 (and thus is itself considered empty). \textbf{DF} sub-volumes are stored similarly. Two integers are stored per voxel---one is a terminating integer, and the other is the L1 distance to the nearest non-empty voxel. The terminating integers and distances are interleaved in memory. \textbf{SVO} sub-volumes are stored as a collection of SVO nodes in the volume buffer. Each SVO node consists of two integers---the first is either a pointer to the beginning of a memory contiguous list of child SVO nodes or is a terminating integer (if the SVO node is an internal or leaf node, respectively), and the second contains valid and leaf masks describing the spatial layout of the octree. The valid mask contains a bit per child describing if that child contains any non-empty voxels, and the leaf mask describes if each child contains a pointer to more SVO nodes, or a terminating integer. SVO nodes contain only a single child pointer because all children nodes are laid out sequentially in memory. We can calculate the pointer to an arbitrary child SVO node using the pointer to the first child node and the valid mask. We store \textbf{SVDAG} sub-volumes similarly to SVO nodes. SVDAG nodes can contain between 1 to 9 integers depending on how many children the node has. The first integer either contains the valid and leaf masks (in internal nodes), or is a terminating integer (in leaf nodes). The next 0 to 8 integers are pointers to children SVDAG nodes. Nodes contain multiple child pointers because there is no guarantee that the children nodes will be laid out sequentially in memory. Figure~\ref{fig:layout} describes these layouts concretely.

\section{The Metaprogramming System} \label{metaprogrammingsystem}

Any combination of base formats can be composed to create a hybrid voxel format. Each format needs construction and ray intersection code for creating and rendering a volume in the format, respectively. We describe a metaprogramming system to automatically generate C++ construction and GLSL ray intersection code for arbitrary hybrid formats. Figure~\ref{fig:metaprogramming_system} shows an example usage of our system, and Figure~\ref{fig:function_proto} shows pseudo-code of the generated construction and intersection functions.

\begin{figure}
\centering

\begin{lstlisting}[language=Python]
construct(lower):
  this_volume = EMPTY, is_empty = True
  for child_lower in morton_next_level_sub_volumes():
    (sub_volume, sub_is_empty) = next_construct(child_lower)
    child_pointer = sub_is_empty ? 0 : 
      write_to_disk(sub_volume)
    this_volume.store(child_pointer, child_lower - lower)
    is_empty = is_empty and sub_is_empty
  return (this_volume, is_empty)
  
intersect(node_id, ray, low):
  for child in ordered_hit_children(node_id, ray):
    if child and next_intersect(child, ray, low + child.pos):
      return True
  return False
\end{lstlisting}

\caption{Pseudo-code for generated level construction and intersection functions. Functions, methods, and fields abstract format specific behavior.}
\label{fig:function_proto}
\end{figure}

\subsection{Hybrid Format Construction}

We build volumes using two tools: the voxelizer, which converts a triangle model into a voxel grid, and the hybrid format construction code, which converts the voxel grid into the target hybrid format. The metaprogramming system generates one construction function per level. We call the smallest coordinates of a voxel in a sub-volume as the "lower indices" of that sub-volume. Each construction function maps the lower indices of a sub-volume to a constructed sub-volume.

For \textbf{Raw} levels, the construction function calls the next level's construction function on each sub-volume. Each non-empty sub-volume from the next level are written to the output file---their positions in the file are stored in a flat array, which is the data for this sub-volume. Once all of the sub-volumes for the next level have been constructed, pushed, and recorded, the flat array is returned. \textbf{DF} levels are constructed similarly---once all of the next level's sub-volumes are recorded, the L1 distance to the nearest non-empty sub-volume is calculated per voxel. For \textbf{SVO} levels, we implement the queue-based algorithm described in \cite{OutOfCoreConstructionVoxel}. The root SVO node is returned as the sub-volume data. \textbf{SVDAG} levels are constructed similarly, except that we maintain a node de-duplication map. When creating SVDAG nodes, we check if each child node is mapped. If one isn't, then we push the child to the file, record its pointer, and add a map entry from the child to its pointer. If one is, then we don't push the child, and use its pointer from the de-duplication map.

The "highest" level is level $1$, and the function for level $N$ calls the function for level $N + 1$. The function for the final level constructs a single voxel by passing the lower indices to the voxelizer and returns the voxel's data. A root construction function is generated, which calls the first construction function and writes the root sub-volume and its pointer to the output file.

\subsection{Hybrid Format Intersection}

We generate one intersection function per level in a format. These functions compute the non-empty sub-volumes hit by a ray; in order of hit time, the next level's function is called on those sub-volumes. Each function maps a sub-volume pointer, a ray, and lower indices to whether the ray hits a single voxel.

We intersect \textbf{Raw} levels using a branchless version of the DDA algorithm \cite{FastVoxelRayTracing}. We intersect \textbf{DF} similarly---the stored L1 distance indicates how many voxels can be marched through before checking occupancy. The intersection functions for \textbf{SVO} and \textbf{SVDAG} levels traverse nodes in pre-order using a per-thread stack. For every hit internal child node, the children nodes are pushed on the stack---the next iteration corresponds to recursing on a child node. For every non-empty hit leaf node, the next level's intersection function is called. 

The metaprogramming system also generates a unit intersection function for single voxels and a root intersection function. The root function first reads the pointer to the root volume from the beginning of the voxel buffer. Next, it checks if the ray intersects the root volume. If so, it calls the first intersection function.

\subsection{Optimizations} \label{optimizations}

Our metaprogramming system implements two optimizations on hybrid formats, one for volume compression and one for ray intersection. The metaprogramming system supports combinations of these optimizations.

By default, each \textbf{SVDAG} sub-volume uses its own de-duplication map. With \textit{whole level de-duplication}, every sub-volume in a \textbf{SVDAG} level shares the same map, improving compression by de-duplicating nodes across sub-volumes.

The per-thread stacks in the intersection functions for the \textbf{SVO} and \textbf{SVDAG} formats are expensive. With \textit{restarting sparse voxel intersection}, we elide this cost by traversing the SVO or SVDAG from the root for every lookup---the downside is the introduction of redundant accesses to the structure.

\section{Out-of-Core Construction} \label{oooconstruction}

\begin{figure}
\centering

\begin{tikzpicture}

\fill[gray!25!white] (0cm,0cm) rectangle (2cm, 2cm);
\fill[gray!10!white] (0cm,1cm) rectangle (1cm, 2cm);

\draw (0.250000, 1.800000) node {$0$};
\draw[blue!70!white, densely dashed, ultra thin] (0.250000, 1.650000) -- (0.750000, 1.650000);
\draw (0.750000, 1.800000) node {$1$};
\draw[blue!70!white, densely dashed, ultra thin] (0.750000, 1.650000) -- (0.250000, 1.150000);
\draw (0.250000, 1.300000) node {$2$};
\draw[blue!70!white, densely dashed, ultra thin] (0.250000, 1.150000) -- (0.750000, 1.150000);
\draw (0.750000, 1.300000) node {$3$};
\draw[blue!70!white, densely dashed, ultra thin] (0.750000, 1.150000) -- (1.250000, 1.650000);
\draw (1.250000, 1.800000) node {$4$};
\draw[blue!70!white, densely dashed, ultra thin] (1.250000, 1.650000) -- (1.750000, 1.650000);
\draw (1.750000, 1.800000) node {$5$};
\draw[blue!70!white, densely dashed, ultra thin] (1.750000, 1.650000) -- (1.250000, 1.150000);
\draw (1.250000, 1.300000) node {$6$};
\draw[blue!70!white, densely dashed, ultra thin] (1.250000, 1.150000) -- (1.750000, 1.150000);
\draw (1.750000, 1.300000) node {$7$};
\draw[blue!70!white, densely dashed, ultra thin] (1.750000, 1.150000) -- (0.250000, 0.650000);
\draw (0.250000, 0.800000) node {$8$};
\draw[blue!70!white, densely dashed, ultra thin] (0.250000, 0.650000) -- (0.750000, 0.650000);
\draw (0.750000, 0.800000) node {$9$};
\draw[blue!70!white, densely dashed, ultra thin] (0.750000, 0.650000) -- (0.250000, 0.150000);
\draw (0.250000, 0.300000) node {$10$};
\draw[blue!70!white, densely dashed, ultra thin] (0.250000, 0.150000) -- (0.750000, 0.150000);
\draw (0.750000, 0.300000) node {$11$};
\draw[blue!70!white, densely dashed, ultra thin] (0.750000, 0.150000) -- (1.250000, 0.650000);
\draw (1.250000, 0.800000) node {$12$};
\draw[blue!70!white, densely dashed, ultra thin] (1.250000, 0.650000) -- (1.750000, 0.650000);
\draw (1.750000, 0.800000) node {$13$};
\draw[blue!70!white, densely dashed, ultra thin] (1.750000, 0.650000) -- (1.250000, 0.150000);
\draw (1.250000, 0.300000) node {$14$};
\draw[blue!70!white, densely dashed, ultra thin] (1.250000, 0.150000) -- (1.750000, 0.150000);
\draw (1.750000, 0.300000) node {$15$};

\end{tikzpicture}

\caption{A depiction of the Morton order of a $4 \times 4$ grid. The upper left $2 \times 2$ sub-grid (light gray) is traversed before the grid cells in the gray region are visited.}
\label{fig:morton}
\end{figure}
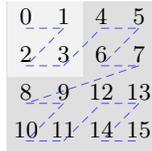

Constructing compressed voxel representations can be challenging, since the input voxel grid may be too large to fit in memory. We implement an out-of-core voxelizer that lazily voxelizes chunks of a triangle model. Like previous work, we convert triangle meshes into voxel volumes in two steps \cite{OutOfCoreConstructionVoxel}. First, we voxelize a triangle model into a raw voxel grid. Second, we compress the raw grid into a hybrid format. Our metaprogramming system implements the second step and our lazy voxelizer implements the first step. Converting triangles into voxels has been well studied previously \cite{VoxelizePoly,HardwareVoxelization,ParallelVoxelization,FastSceneVoxelization}. The voxelizer lazily creates a new chunk of voxels when the construction code requests a voxel in a new chunk.

Morton order is a linear ordering of points in higher dimensional spaces. Baert et al. \cite{OutOfCoreConstructionVoxel} observe that Morton order corresponds to post-order traversals of octrees and uses that fact to construct SVOs out-of-core. We arrange \textit{all} accesses to the voxelizer in Morton order. In particular, we modify the \textbf{Raw} and \textbf{DF} construction functions to iterate over child sub-volumes in Morton order. By iterating over single voxels in Morton order, the voxelizer can guarantee that all of the individual voxels in a chunk will be visited before other chunks. The only requirement on the used format is that each level but the highest have cubic power-of-two dimensions, since Morton order is hierarchical with respect to sub-volumes of power-of-two size. Figure~\ref{fig:morton} visualizes a Morton order traversal.

\begin{figure*}[ht]
    \centering
    \begin{subfigure}{0.24\textwidth}
    \centering
    \includegraphics[height=1.6cm]{san-miguel.png}
    \end{subfigure}
    \begin{subfigure}{0.24\textwidth}
    \centering
    \includegraphics[height=1.6cm]{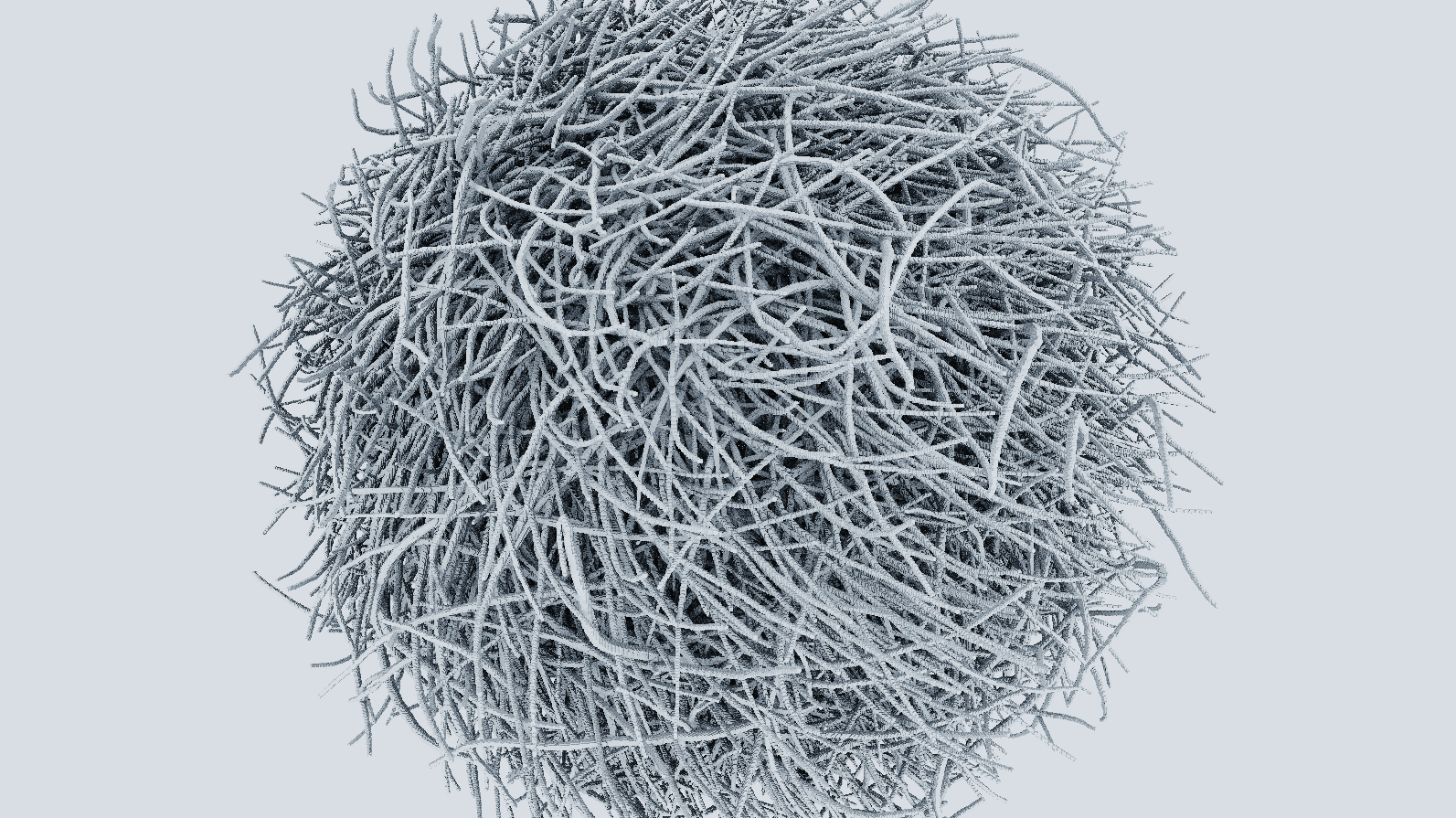}
    \end{subfigure}
    \begin{subfigure}{0.24\textwidth}
    \centering
    \includegraphics[height=1.6cm]{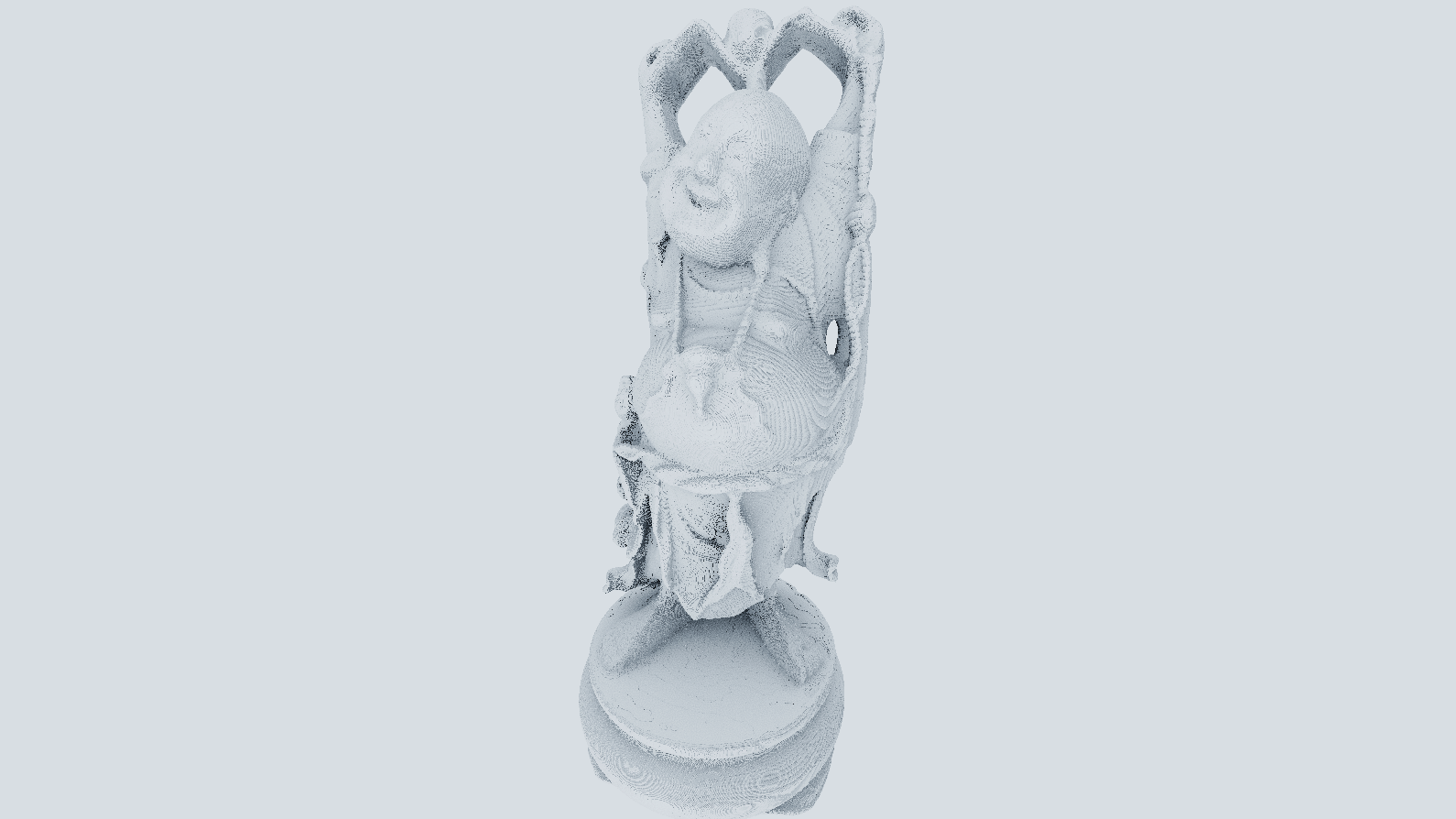}
    \end{subfigure}
    \begin{subfigure}{0.24\textwidth}
    \centering
    \includegraphics[height=1.6cm]{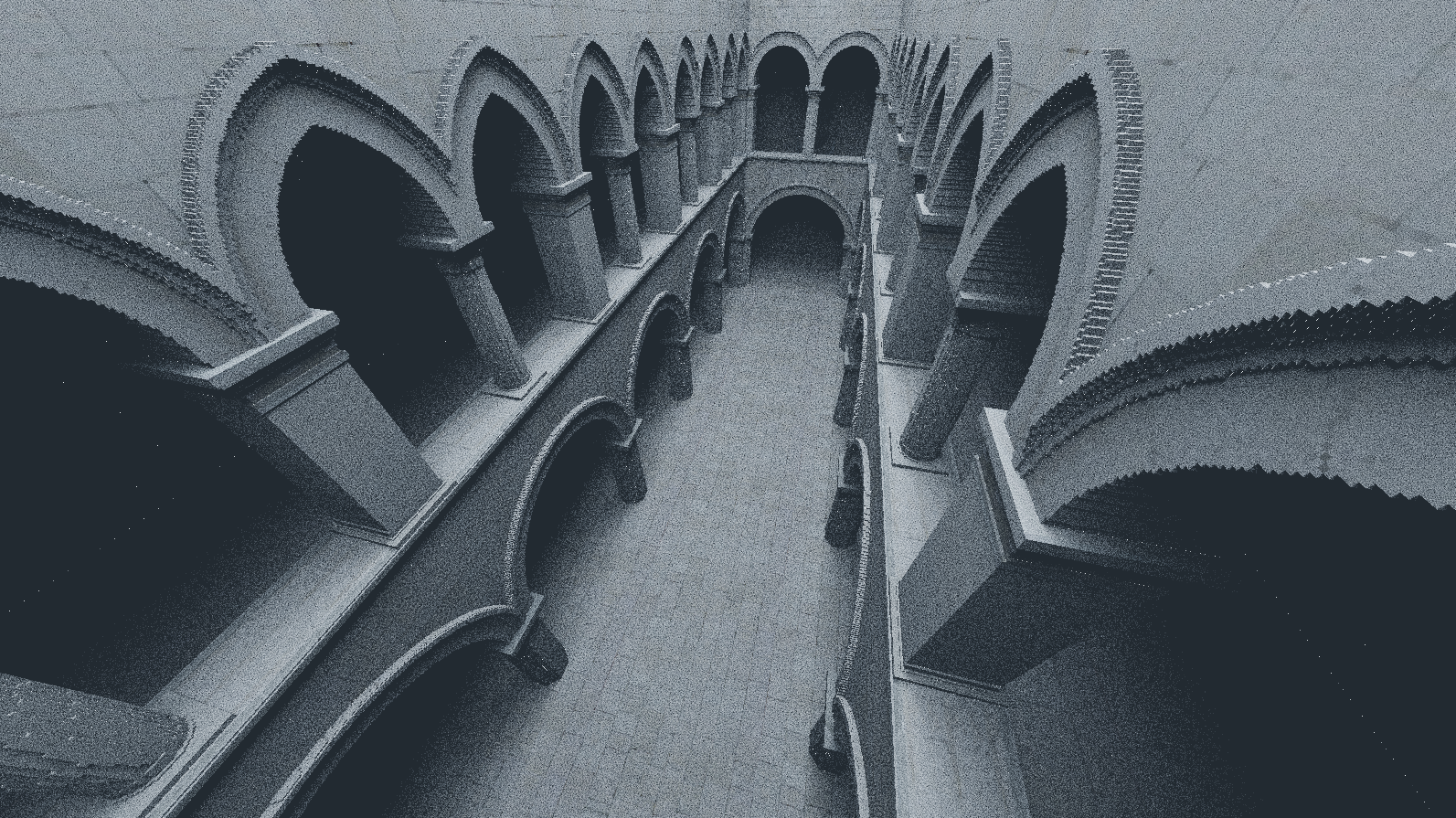}
    \end{subfigure}
    \caption{Screenshots of the tested models from the camera positions used in experiments. The experiments are run without shading from bounces.}
    \label{fig:pretty}
\end{figure*}

\section{Evaluation} \label{evaluation}

We evaluate 40 formats (Table~\ref{tab:2048}) on four models: San Miguel, Hairball, Buddha, and Sponza \cite{model_archive,HairballPaper,BuddhaPaper}. We observe new Pareto optimal formats, shown in Figures~\ref{fig:figure_1} and \ref{fig:figure_2}. We evaluate our two optimizations, realizing up to 2.1x increased rendering performance and 4.7x smaller storage sizes.

We implement the metaprogramming system and out-of-core voxelizer as tools for the Illinois Voxel Sandbox, a general purpose voxel volume rendering engine written in C++ and using the Vulkan graphics API. We use custom intersection and closest-hit shaders to integrate hybrid voxel formats. Our implementation is open source\footnote{\url{https://github.com/RArbore/illinois-voxel-sandbox}}. All of our experiments were run on a machine with an AMD Ryzen 7 2700X CPU, 64 GiB of main memory, and a NVIDIA RTX 2060 Super GPU. We measure frame time using an in-engine counter.

\begin{table}
  \centering
  \begin{tabular}{c c c c}
    \toprule
    Label & $2048^3$ Formats & Label & $512^3$ Formats \\
    \midrule
1 & D($4^3$, 6) D($3^3$, 6) G(4) & 21 & D($3^3$, 6) D($3^3$, 6) G(3) \\
2 & D($4^3$, 6) R($3^3$) G(4) & 22 & D($3^3$, 6) R($3^3$) G(3) \\
3 & R($4^3$) R($3^3$) G(4) & 23 & R($3^3$) R($3^3$) G(3) \\ 
4 & R($4^3$) S(3) G(4) & 24 & R($3^3$) R($3^3$) R($3^3$)  \\ 
5 & R($4^3$) R($4^3$) R($3^3$) & 25 & R($4^3$) R($1^3$) R($4^3$)  \\ 
6 & D($4^3$, 6) S(7) & 26 & D($4^3$, 6) S(5)  \\ 
7 & D($4^3$, 6) G(7) & 27 & D($4^3$, 6) G(5)  \\ 
8 & D($6^3$, 6) S(5) & 28 & R($4^3$) S(5)  \\ 
9 & D($6^3$, 6) G(5) & 29 & R($4^3$) G(5)  \\ 
10 & R($4^3$) S(7) & 30 & R($2^3$) G(7)  \\ 
11 & R($4^3$) G(7) & 31 & R($7^3$) G(2)  \\ 
12 & R($6^3$) S(5) & 32 & S(5) R($4^3$)  \\ 
13 & R($6^3$) G(5) & 33 & G(5) R($4^3$)  \\ 
14 & R($3^3$) G(8) & 34 & D($5^3$, 6) D($4^3$, 6)  \\ 
15 & R($8^3$) G(3) & 35 & D($5^3$, 6) R($4^3$)  \\ 
16 & S(7) G(4) & 36 & R($5^3$) R($4^3$)  \\ 
17 & S(5) G(6) & 37 & R($9^3$)  \\ 
18 & S(3) G(8) & 38 & D($9^3$, 6)  \\ 
19 & S(11) & 39 & S(9)  \\ 
20 & G(11) & 40 & G(9)  \\ 
    \bottomrule
  \end{tabular}
  \caption{Hybrid formats we evaluate. Formats 1-20 represent a $2048^3$ grid, and formats 21-40 represent a $512^3$ grid.}
  \label{tab:2048}
\end{table}

\subsection{Evaluating Hybrid Formats}

\begin{figure}
\centering
    \includegraphics[width=0.45\textwidth]{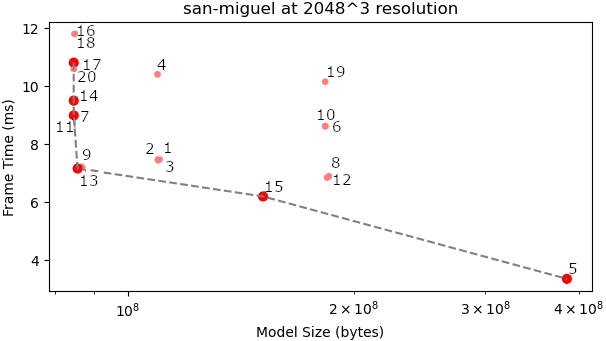}
    \includegraphics[width=0.45\textwidth]{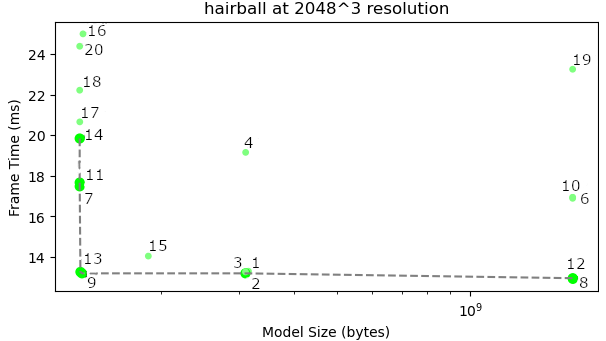}
    \includegraphics[width=0.45\textwidth]{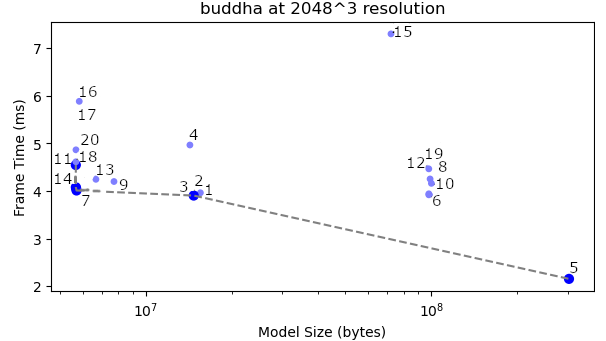}
    \includegraphics[width=0.45\textwidth]{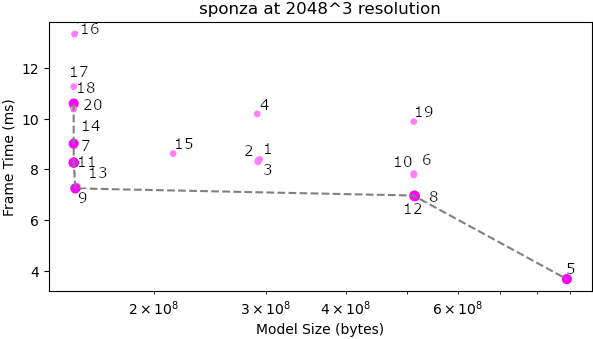}
    \caption{Graphs of rendering performance vs. storage size for 4 models and 20 $2048^3$ formats. The Pareto frontier is emphasized. Models too large to fit on the GPU are omitted. Lower on both axes is better.}
    \label{fig:figure_1}
\end{figure}
\begin{figure}
\centering
    \includegraphics[width=0.45\textwidth]{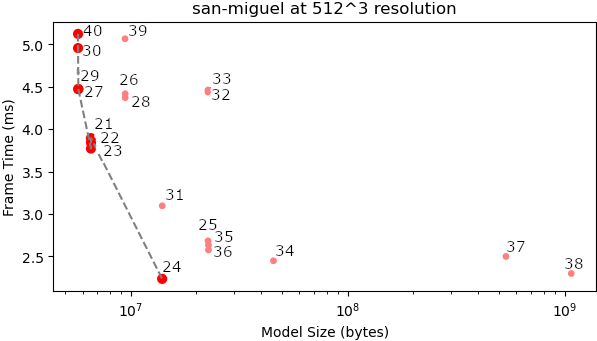}
    \includegraphics[width=0.45\textwidth]{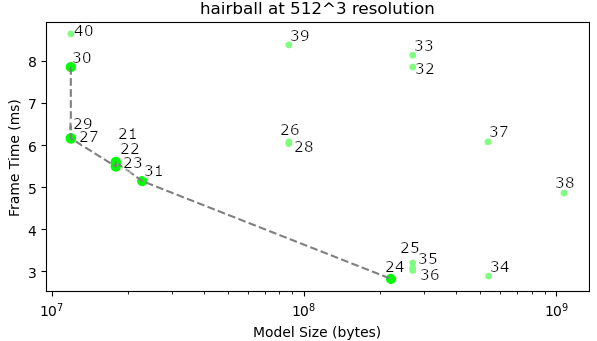}
    \includegraphics[width=0.45\textwidth]{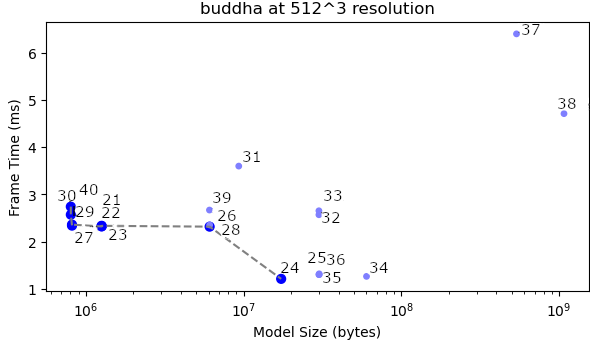}
    \includegraphics[width=0.45\textwidth]{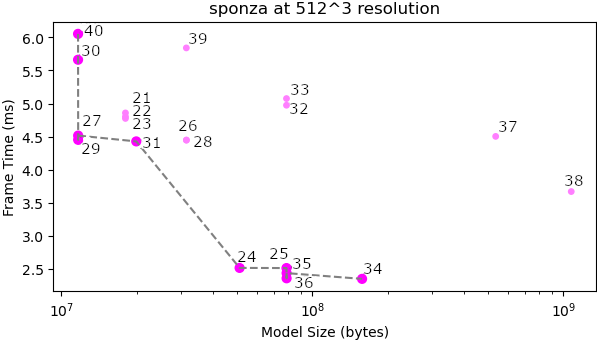}
    \caption{Graphs of rendering performance vs. storage size for 4 models and 20 $512^3$ formats. The Pareto frontier is emphasized. Lower on both axes is better.}
    \label{fig:figure_2}
\end{figure}

We selected formats using \textbf{Raw} and \textbf{DF} upper levels and \textbf{SVO} and \textbf{SVDAG} lower levels---the number of sub-volumes grows exponentially in lower levels, so less compact lower levels explode storage size. We constructed and rendered each model in each format with whole level de-duplication enabled. Figures~\ref{fig:figure_1} and \ref{fig:figure_2} show the trade-off between memory usage and rendering performance.

Two $2048^3$ formats always lie on the Pareto frontier: R($3^3$) G($8$) and R($4^3$) G($7$). Both of these formats balance culling space, de-duplicating homogeneous regions, and intersection complexity. For three models, the D($6^3$, $6$) G($5$) and R($6^3$) G($5$) formats achieve better rendering performance at competitive model sizes. For the Buddha model, these formats are slower, likely because Buddha contains lots of empty space, which isn't accelerated by uncompressed voxel grids. The tested base formats, S($11$) and G($11$), show worse rendering performance than the hybrid formats. G($11$) produces very compact models, while S($11$) provides poor compression in comparison. We didn't test \textbf{Raw} or \textbf{DF} base formats, since the models would be too large to render.

Four $512^3$ formats always lie on the Pareto frontier: R($3^3$) R($3^3$) R($3^3$), D($4^3$, 6) G($5$), R($4^3$) G($5$), and R($2^3$) G($7$). R($3^3$) R($3^3$) R($3^3$) achieves the best intersection performance for three of the models, but is an order of magnitude larger than the smallest formats. Both D($4^3$, $6$) G($5$) and R($4^3$) G($5$) achieve nearly the best compression, while having good intersection speed.

We observe several patterns in the performance and size of hybrid voxel formats. Although \textbf{SVO} levels can theoretically be smaller than \textbf{SVDAG} levels due to having smaller nodes, we observe that \textbf{SVDAG} levels are often more compact in practice. We observe that using voxel grids (either \textbf{Raw} or \textbf{DF} levels) that are too granular for models with lots of empty space can be detrimental to performance, since voxel grids cannot compress empty space within a single voxel grid level. We suggest a good default format for most models is a format of the form R($N^3$) G($M$)---formats of this form consistently perform well in our experiments. However, we emphasize that the best format for a particular model depends on the sparsity and homogeneity of the volume, the resolution, and the desired trade-off between intersection performance and storage cost.

\subsection{Evaluating Optimizations}

\begin{figure}
\centering
\begin{minipage}{.46\textwidth}
  \centering
    \includegraphics[width=\textwidth]{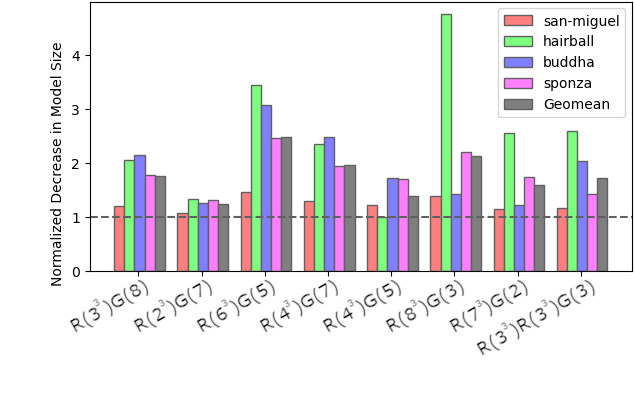}
    \captionof{figure}{Normalized decrease in model size after \texttt{-whole-level-dedup}.}
    \label{fig:whole-level-dedup}
\end{minipage}%
\hspace{0.01\textwidth}
\begin{minipage}{.46\textwidth}
  \centering
    \includegraphics[width=\textwidth]{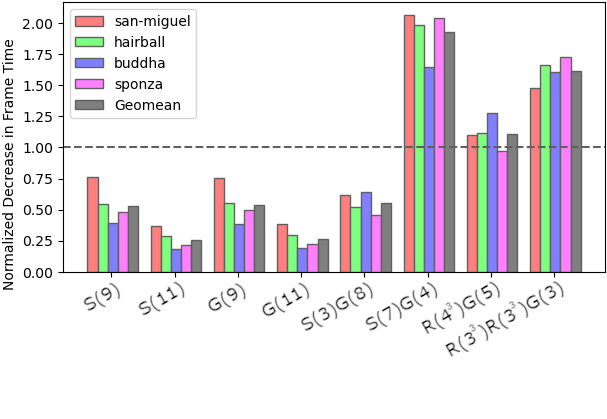}
    \captionof{figure}{Normalized decrease in frame time after \texttt{-restart-sv}.}
    \label{fig:restart-sv}
\end{minipage}
\end{figure}

We evaluate whole level de-duplication on formats with at least one \textbf{SVDAG} lower level. Shown in Figure~\ref{fig:whole-level-dedup}, we observe consistent reductions in model size---performing node de-duplication across sub-volumes doesn't add any storage overhead. The highest compression factor achieved is 4.74x, when storing the Hairball model in the R($8^3$) G($3$) format---this is because all voxels in Hairball are the same color. Shown in Figure~\ref{fig:restart-sv}, restarting sparse voxel intersection provides rendering performance improvements on 3 out of 8 tested formats. These formats all have a shallow \textbf{SVDAG} lowest level, meaning the intersection code re-traverses fewer SVDAG nodes per lookup. The G($11$) format performs poorly with restarting intersection because the intersection code must traverse every layer in the SVDAG for every access to an individual voxel.

\subsection{Hybrid Format Construction}

\begin{figure}
    \centering
    \includegraphics[width=0.47\textwidth]{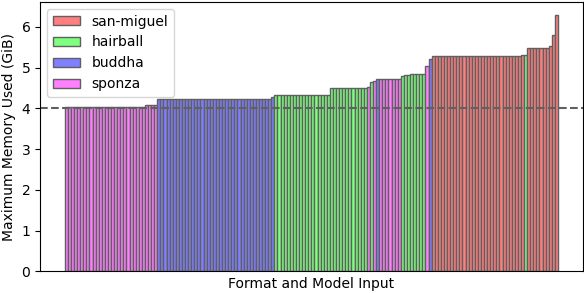}
    \caption{Maximum memory used per model, per format. Bars are ordered by maximum memory used. Voxelization uses a 4 GiB scratchpad.}
    \label{fig:max_mem}
\end{figure}

We show that our method can construct compressed voxel models in an out-of-core fashion by measuring the maximum amount of memory used while constructing each model in each format---these amounts are shown in Figure~\ref{fig:max_mem}. We set the voxel chunking size to 4 GiB. The variance across models mostly corresponds to the size of each models' triangle mesh (we don't pre-chunk triangles to perform out-of-core construction for the input triangle model). The D($9^3$, $6$) format requires the most memory across all models because the entire $512^3$ grid is needed in memory to compute the distance field. However, the amount of memory used is always significantly lower than the 32 GiB required to store an intermediate $2048^3$ grid.

\section{Future Work}
We have presented a formulation of hybrid voxel formats and a system to evaluate these formats. Our system is flexible enough to support alternative implementations of various aspects of hybrid formats, and supports out-of-core construction. We present an extensive evaluation of hybrid formats using our system, providing insight into many of the trade-offs involved in selecting a voxel format. 

Our experiments were limited by the set of manually-selected formats---there are potentially model-specific outlier hybrid formats that optimally take advantage of each model's sparsity and homogeneity patterns. An avenue for future work could be automatically exploring the search space of hybrid formats \cite{Spire}. It could also be fruitful to employ deep learning models to generate hybrid format proposals for different voxel models. This work describes an infrastructure capable of generating the dataset for training such a model.

\bibliographystyle{splncs04}
\bibliography{main}
\end{document}